\documentclass[conference]{IEEEtran}
\IEEEoverridecommandlockouts
\usepackage{amsmath,amssymb,amsfonts}
\usepackage{algorithmic}
\usepackage{graphicx}
\usepackage{textcomp}
\usepackage{bmpsize}
\usepackage{xcolor}
\usepackage{subcaption}
\usepackage{booktabs}

\usepackage{enumitem}
\usepackage{threeparttable}
\usepackage{multirow}

\usepackage{tikz}

\usepackage{siunitx} 
\usepackage[
 backend=biber
,style=ieee
,minbibnames=1
,maxbibnames=5
,maxcitenames=1
,mincitenames=1
,eprint=true
,doi=true
,isbn=false
,url=true
]{biblatex}
\bibliography{ref} 
\AtBeginBibliography{\footnotesize}

\DeclareSourcemap{
  \maps{
    \map{
      \step[fieldset=month, null]
      \step[fieldset=address, null]
      \step[fieldset=location, null]
      \step[fieldset=publisher, null]
      \step[fieldset=url, null]
      \step[fieldset=isbn, null]
      \step[fieldset=series, null]
      \step[fieldset=editor, null]
    }
  }
}

\usepackage{cleveref}
\Crefname{figure}{Fig.}{Figs.}

\usepackage{lipsum}

\def\BibTeX{{\rm B\kern-.05em{\sc i\kern-.025em b}\kern-.08em
    T\kern-.1667em\lower.7ex\hbox{E}\kern-.125emX}}

\usepackage{flushend}

\begin{document}

\title{Binary VPN Traffic Detection Using Wavelet Features and Machine Learning}

\author{
    \IEEEauthorblockN{Yasameen Sajid Razooqi\IEEEauthorrefmark{1} and Adrian Pekar\IEEEauthorrefmark{1}\IEEEauthorrefmark{2}\IEEEauthorrefmark{3}
    }
    \IEEEauthorblockA{
        \IEEEauthorrefmark{1}
        Budapest University of Technology and Economics, M\H{u}egyetem rkp. 3., H-1111 Budapest, Hungary.\\
        \IEEEauthorrefmark{2}HUN-REN-BME Information Systems Research Group, Magyar Tud\'{o}sok krt. 2, 1117 Budapest, Hungary.\\
        \IEEEauthorrefmark{3}CUJO LLC, Budapest, Hungary.\\
        E-mail: rsajid@hit.bme.hu, apekar@hit.bme.hu
    }
}

\maketitle

\begin{abstract}
Encrypted traffic classification faces growing challenges as encryption renders traditional deep packet inspection ineffective. This study addresses binary VPN detection, distinguishing VPN-encrypted from non-VPN traffic using wavelet transform-based features across multiple machine learning models. Unlike previous studies focused on application-level classification within encrypted traffic, we specifically evaluate the fundamental task of VPN identification regardless of application type. We analyze the impact of wavelet decomposition levels and dataset filtering on classification performance across significantly imbalanced data, where filtering reduces some traffic categories by up to 95\%. Our results demonstrate that Random Forest (RF) achieves superior performance with an F1-score of 99\%, maintaining robust accuracy even after significant dataset filtering. Neural Networks (NN) show comparable effectiveness with an F1-score of 98\% when trained on wavelet level 12, while Support Vector Machines (SVM) exhibit notable sensitivity to dataset reduction, with F1-scores dropping from 90\% to 85\% after filtering. Comparing wavelet decomposition at levels 5 and 12, we observe improved classification performance at level 12, particularly for variable traffic types, though the marginal gains may not justify the additional computational overhead. These findings establish RF as the most reliable model for VPN traffic classification while highlighting key performance tradeoffs in feature extraction and preprocessing.
\end{abstract}


\begin{tikzpicture}[remember picture,overlay]
\node[anchor=north, align=center, text=black, font=\small\itshape, yshift=-.6cm] at (current page.north) {This version of the paper was accepted for presentation at the 2025 International Conference on Software,\\ Telecommunications, and Computer Networks (SoftCOM 2025)};
\end{tikzpicture}

\section{Introduction}

The widespread adoption of network encryption poses a challenge for traffic classification, as traditional deep packet inspection and port-based methods are no longer effective~\cite{CiscoVNI2019,GoogleTransparency2020}. VPNs are among the most commonly used encryption tools, with nearly one-third of internet users employing VPN services to protect their online activity~\cite{Top10VPN2020}. While this enhances privacy and security, it also complicates traffic analysis for network management, intrusion detection, and policy enforcement.

Recent research has explored machine learning (ML) for encrypted traffic classification. The approaches in this field can be broadly categorized into three main groups. First, traditional feature-based methods have shown promise, with techniques like Sample Entropy Fingerprint and TCP characteristics heuristics achieving high accuracy in VPN detection~\cite{9353766,hanlon2024detecting}. Second, deep learning approaches have emerged as powerful tools, with innovations like FlowPics for visual representation of flow features~\cite{10724208} and advanced architectures combining CNN with attention mechanisms~\cite{Chai2024,Seydali2024}. The third category explores novel modeling paradigms, including multi-task learning with DistilBERT~\cite{Park2024} and text-to-text transformation using large language models~\cite{Luo2024}.

Notably, \citeauthor{10044382}\cite{10044382} introduced an ML framework that combines wavelet-based features with uncertainty quantification for application-level classification across VPN and non-VPN traffic. However, the related but distinct problem of binary VPN detection---determining whether traffic is VPN-encrypted regardless of application---has received less attention and presents unique value for network policy enforcement and security monitoring. While wavelet transforms have proven effective in network anomaly detection\cite{4023145,4198844}, their potential for binary VPN classification remains unexplored. Building on VNAT's approach, our study applies wavelet-based feature extraction specifically for binary VPN detection. While both studies use wavelet decomposition, our work differs from VNAT in three key aspects: objectives (application classification vs. binary VPN detection), methodology (we systematically evaluate decomposition depths J=12 and J=5, finding J=12 superior despite diminishing returns), and architecture (demonstrating that RF achieves comparable performance to VNAT's specialized Prototypical Network).

In this work, we focus on \textit{binary VPN detection} and re-evaluate the effectiveness of \textit{wavelet transform-based features} for distinguishing VPN from non-VPN traffic. We assess the performance of multiple ML models, including Random Forest (RF), Neural Networks (NN), and Support Vector Machines (SVM), and analyze the impact of dataset filtering, a common preprocessing step that significantly affects classification performance and data distribution. Our analysis deliberately includes both filtered and unfiltered datasets to evaluate the robustness of models across varying conditions, even with the highly imbalanced class distributions present in real-world traffic. Unlike prior work that primarily targets application classification, our analysis provides insights into the robustness of different ML models under specific data processing variations: wavelet decomposition depth (levels 5 and 12) and flow-length filtering.

Our results show that RF achieves the highest accuracy, reaching 99\% with wavelet level 12 and maintaining strong performance (98\%) even after dataset filtering. NNs also perform well, with accuracy dropping slightly from 98\% to 96\% after filtering. SVMs experience the most significant decline, with accuracy decreasing from 90\% to 85\% (wavelet level 12) and 88\% to 83\% (wavelet level 5) after filtering. Detailed misclassification analysis of the best-performing RF model reveals that Chat traffic exhibits the highest error rate, with frequent confusion between VPN and non-VPN traffic, even after filtering. In contrast, VoIP traffic is classified with perfect accuracy, showing no misclassification in either VPN or non-VPN scenarios. While deeper wavelet decomposition (level 12 versus level 5) improves classification accuracy, particularly for variable traffic types, the modest performance gains may not justify the additional computational overhead. These findings emphasize the importance of using robust models like RF for VPN detection, especially when dataset size is limited.

The remainder of this paper is structured as follows: 
\Cref{sec:method} describes our methodology, including dataset preprocessing and model selection. 
\Cref{sec:res} presents and analyzes our results. 
Finally, \Cref{sec:conc} concludes this paper. 

\begin{table*}[]
    \footnotesize
    \centering
    \caption{Comparison of Flow Counts with Different Active Timeouts}
    \begin{tabular}{lrrrrrrrrrrr}
        \toprule
         & \multicolumn{3}{c}{Complete Flows} & & \multicolumn{5}{c}{Act. Timeout 41} & \multicolumn{1}{c}{Act. Timeout 40.96} \\
        \cmidrule(r){2-5} \cmidrule(r){6-10} \cmidrule(r){11-11}
        Category & non-VPN & VPN & $\sum$ & Tab. 3~\cite{10044382} & non-VPN & VPN & $\sum$ & Filtered & Reduction (\%) & Sec.~IV.A~\cite{10044382} \\
        \midrule
        Chat & 1244 & 57 & 1301 & 1301 & 5928 & 5751 & 11679 & 10259 & 12.16 & 10498 \\
        Com. \& Cont. & 13591 & 8 & 13599 & 13599 & 15961 & 1624 & 17585 & 1470 & 91.64 & 1675 \\
        File Transfer & 16420 & 10 & 16430 & 16430 & 17588 & 378 & 17966 & 805 & 95.52 & 851 \\
        Streaming & 1759 & 5 & 1764 & 1764 & 3334 & 160 & 3494 & 1841 & 47.31 & 1827 \\
        VoIP & 318 & 299 & 617 & 617 & 786 & 437 & 1223 & 242 & 80.21 & 243 \\
        \bottomrule
    \end{tabular}
\label{tbl:DS}
\end{table*}

\section{Methodology}
\label{sec:method}


\subsection{Dataset Description}
\label{sec:DS}

We use the VPN/non-VPN Network Application Traffic (VNAT)~\cite{10044382} dataset for our classification task. The dataset comprises 165 pcap files (82 VPN and 83 non-VPN traffic traces) with a total size of 36.1 GB. All network traffic was captured using \texttt{tcpdump} and stored in PCAP format, encompassing both VPN-encrypted and unencrypted flows across 33,711 connections and approximately 272 hours of capture time.

The dataset covers traffic from 10 applications categorized into 5 main groups: Streaming, VoIP, Chat, Command \& Control, and File Transfer. These categories exhibit notable variations in their distribution across different metrics. For instance, while File Transfer represents 90\% of the dataset's volume, it accounts for only 40\% of connections and 5\% of the total capture time. Conversely, Command \& Control traffic, despite comprising merely 2\% of the dataset's size, represents 40\% of all connections and 47\% of the total capture duration.

\subsection{Flow Processing and Analysis}

We utilized NFStream~\cite{Aouini2022}, a flexible network data analysis framework, to process the dataset's PCAP files. While VNAT provides both traffic traces in PCAP format and pre-extracted statistical summaries in HDF (h5) format, we chose NFStream for its ability to seamlessly integrate plugins, particularly for calculating the Wavelet Features discussed in \Cref{sec:WF}.

To ensure consistency with the methodology described in~\cite{10044382}, we extracted flow records under two conditions: first, without applying any flow timeout to capture complete flow records, and second, with an active timeout of 41 seconds, approximating the original study's segmentation of flows into 40.96-second intervals. In the latter case, packets were assigned to 0.01-second time bins, and intervals containing fewer than 20 packets were discarded. This dual approach enables us to evaluate the impact of flow segmentation and filtering on classification performance.

\Cref{tbl:DS} presents the comparative flow counts, revealing that while complete flow counts match exactly with the original study, the application of timeout parameters leads to notable variations across traffic categories:
\begin{itemize}
\item \textit{Chat}: 1301 complete flows in both studies; our timeout yielded 10259 flows versus original 10498 (12.16\% reduction after filtering)
\item \textit{Command \& Control}: 13599 complete flows; timeout processing produced 1470 flows versus 1675 (91.64\% reduction after filtering)
\item \textit{File Transfer}: 16430 complete flows; resulted in 805 flows versus 851 (95.52\% reduction after filtering)
\item \textit{Streaming}: 1764 complete flows; yielded 1841 flows versus 1827 (47.31\% reduction after filtering)
\item \textit{VoIP}: 617 complete flows; produced 242 flows versus 243 (80.21\% reduction after filtering)
\end{itemize}

These substantial differences in flow counts, despite similar filtering criteria, demonstrate the sensitivity of flow segmentation techniques to minor timing parameter variations (41s vs. 40.96s), as NFStream only supports integer timeout values for performance optimization. 
 
The reduction in flow counts is particularly pronounced in Command \& Control (91.64\%), File Transfer (95.52\%), and VoIP (80.21\%) traffic. This dramatic reduction highlights the effect of flow segmentation on dataset composition and raises potential concerns regarding class imbalance.

Beyond differences in flow segmentation, we identified two key methodological variations that distinguish our processing from the original VNAT dataset. First, NFStream processes only the first fragment (offset zero) of fragmented IP packets, as it requires flow headers for proper stream reconstruction. This means that while our flow segmentation remains comparable, NFStream does not reconstruct fragmented packets to obtain their true size, whereas the original study performed full reassembly. This methodological difference primarily affects large UDP-based transmissions where fragmentation occurs more frequently.

Second, we observed discrepancies in how transport layer sizes were computed. The original study inconsistently applied UDP header exclusions while including TCP headers in size calculations. NFStream, in contrast, consistently accounts for TCP and UDP payload size differences, ensuring uniformity in flow size calculations. While these differences do not directly impact the extracted features, they affect dataset comparability and highlight the sensitivity of flow statistics to implementation details.

This filtering has a non-uniform effect on VPN and non-VPN traffic distributions, altering the class balance within categories and creating a more challenging classification scenario that better reflects real-world deployment conditions.

\subsection{Wavelet Features}
\label{sec:WF}

Wavelet transform is a time-frequency analysis method that decomposes signals into components at different scales, making it suitable for analyzing non-stationary traffic patterns. Unlike Fourier transforms, wavelets preserve both time and frequency information, capturing traffic bursts and periodic behaviors effectively~\cite{Ahad2016}.

The Discrete Wavelet Transform (DWT) decomposes a signal into approximation coefficients (low-frequency components) and detail coefficients (high-frequency components) at multiple levels. For a network flow, the signal is the sequence of packet sizes over time~\cite{91217}. The approximation coefficients are computed as:

{\footnotesize
\begin{equation}
A_j[k] = \sum_{n} x[n] \cdot g(2k - n)
\end{equation}
}

where $x[n]$ is the input signal, $g[n]$ represents the low-pass filter coefficients, and $2k$ is the downsampling factor. The detail coefficients are computed as:

{\footnotesize
\begin{equation}
D_j[k] = \sum_{n} x[n] \cdot h(2k - n)
\end{equation}
}

where $h[n]$ represents the high-pass filter coefficients.

From these coefficients, we extract four key metrics that characterize different aspects of network traffic:

\begin{enumerate}
\item \textit{Relative Wavelet Energy Vector ($E_j$)} represents the energy distribution across different decomposition levels:
{\footnotesize
\begin{equation}
E_j = \frac{\sum_{k} C_j[k]^2}{\sum_{j}\sum_{k} C_j[k]^2} \times 100\%
\end{equation}
}
\item \textit{Absolute Mean of Coefficients ($M_j$)} quantifies the average magnitude at level $j$ ~\cite{753747}:
{\footnotesize
\begin{equation}
M_j = \frac{1}{N_j} \sum_{k} |C_j[k]|
\end{equation}
}

\item \textit{Standard Deviation of Coefficients ($\sigma_j$)} measures coefficient spread~\cite{7566534}:
{\footnotesize
\begin{equation}
\sigma_j = \sqrt{\frac{1}{N_j} \sum_{k} \left(C_j[k] - \bar{C}_j\right)^2}
\end{equation}
}

\item \textit{Shannon Entropy ($H_j$)} quantifies the information content~\cite{Shannon1948}:
{\footnotesize
\begin{equation}
H_j = -\sum_{k} P_j[k] \log_2 P_j[k]
\end{equation}
}
where $P_j[k] = \frac{|C_j[k]|}{\sum_{k} |C_j[k]|}$.
\end{enumerate}

The optimal number of decomposition levels is determined as $J = \lfloor \log_2(n) \rfloor$, where $n$ is the sequence length. This provides a mathematical basis for our choice of decomposition levels in feature extraction.

\subsection{Feature Extraction}

For our classification task, we extract wavelet-based features to capture the temporal characteristics of packet size distributions in network flows. We apply DWT to transport-layer packet size sequences in both forward and backward directions. Following \cite{10044382}, we remove flows with fewer than 20 packets. Our choice of decomposition level J=5 is supported by the packet count distribution in our filtered dataset, as shown in \Cref{tbl:PC}. With the 25th percentile at 56 packets, J=5 (scale of $2^{5} = 32$ according to the optimal decomposition level formula $J = \lfloor \log_2(n) \rfloor$) ensures effective capture of packet variations across the majority of flows. We also evaluate J=12 to enable direct comparison with \cite{10044382}, though at this level (scale of $2^{12} = 4096$), the decomposition scale significantly exceeds typical flow lengths in their dataset. The highly skewed distribution (mean $=2,599$, median $=63$) indicates that while some flows contain up to 857,900 packets, such extensive flows are rare outliers. This distribution suggests that while J=12 might capture finer-grained variations in exceptionally long flows, it might provide diminishing returns for typical traffic patterns.

\begin{table}[h]
    \centering
    \caption{Packet Count Statistics Before and After Filtering}
    \label{tab:packet_stats}
    \begin{tabular}{lcc}
        \hline
        \textbf{Statistic} & \textbf{Before Filtering} & \textbf{After Filtering} \\
        \hline
        Count & 51947 & 14617 \\
        Mean & 733.50 & 2599.00 \\
        Standard Deviation & 12444.49 & 23357.13 \\
        Minimum & 1 & 20 \\
        25th Percentile & 2 & 56 \\
        Median & 2 & 63 \\
        75th Percentile & 36 & 73 \\
        Maximum & 857900 & 857900 \\
        \hline
    \end{tabular}
\label{tbl:PC}    
\end{table}

From each decomposition, we compute four key wavelet-based metrics using both the approximation coefficients and detail coefficients: relative wavelet energy, absolute mean, standard deviation, and Shannon entropy. These metrics quantify the distribution of energy across frequency components, capturing structural properties of packet sequences.

At each decomposition level \(j\), these four metrics are computed separately for the corresponding detail coefficients \(D_j\). Additionally, at the final decomposition level \(J\), they are also computed for the approximation coefficients \(A_J\), leading to \( 4J \) features from the \(J\) sets of detail coefficients and \( 4 \) features from the final approximation coefficients.

Since we extract these features separately for both the forward and backward packet sequences, the total number of wavelet-based features is given by \(F_{wavelet} = 8J + 8\), where \(8J\) accounts for the features extracted from all \(J\) levels of detail coefficients, while the additional \(8\) corresponds to the features from the final approximation coefficients. This results in 48 features for \(J=5\) and 104 features for \(J=12\), ensuring a balance between feature diversity and computational efficiency.

\subsection{Classification Models}

To classify VPN and non-VPN traffic, we employ three machine learning models: RF, NN, and SVM. Each model is evaluated on both full and filtered datasets, with an 80/20 train-test split and fixed random seed to ensure reproducibility.

\subsubsection{Random Forest} RF is implemented as an ensemble-based classifier, leveraging decision tree aggregation to improve generalization. 
Default hyperparameters were used, as preliminary evaluation showed competitive performance across the dataset without requiring extensive optimization.

\subsubsection{Neural Networks}

NN model consists of three fully connected layers: an input layer with 64 neurons, a hidden layer with 32 neurons using ReLU activation, and an output layer with a sigmoid activation function for binary classification. The model is trained using the Adam optimizer with binary cross-entropy loss for 20 epochs. 

\subsubsection{Support Vector Machine}

SVM is trained using a linear kernel and optimized with \texttt{LinearSVC}, configured with a maximum iteration limit of 10,000 to ensure convergence. We apply feature scaling prior to training to stabilize optimization.

The chosen model configurations demonstrated strong classification performance across the dataset without requiring extensive hyperparameter tuning. Given their efficiency and stability, they provide the groundwork for future research employing hyperparameter optimization and alternative models.

\subsection{Performance Metrics}

We evaluate model performance using standard classification metrics: accuracy, precision, recall, and F1-score. Additionally, we analyze confusion matrices and misclassified samples to assess model behavior across different traffic categories. Misclassification patterns provide insights into category-specific classification challenges and help identify traffic types that are particularly difficult to distinguish under VPN obfuscation.

Beyond classification metrics, we investigate the effect of filtering on model performance. Since filtering reduces dataset size by discarding flows with fewer than 20 packets, it may introduce biases or reduce model generalizability. We deliberately evaluate performance on both filtered and unfiltered datasets to determine how robust each model is to variations in flow length, an important consideration for real-world deployment where flows of all lengths will be encountered. 
\Cref{sec:res} discusses the impact of filtering on different models.

Analyzing both filtered and unfiltered datasets, despite the predominance of very short flows (1-2 packets) in the latter, serves two important purposes: first, it establishes a performance baseline for comparison across preprocessing techniques; second, it evaluates the models' capability to extract meaningful signal from minimal data, which is essential in scenarios where early traffic detection is required before sufficient packets have been observed.

\section{Results}
\label{sec:res}

\begin{figure}[!htp]
    \centering
    \includegraphics[width=1\linewidth,trim=0 0 0 .9cm,clip]{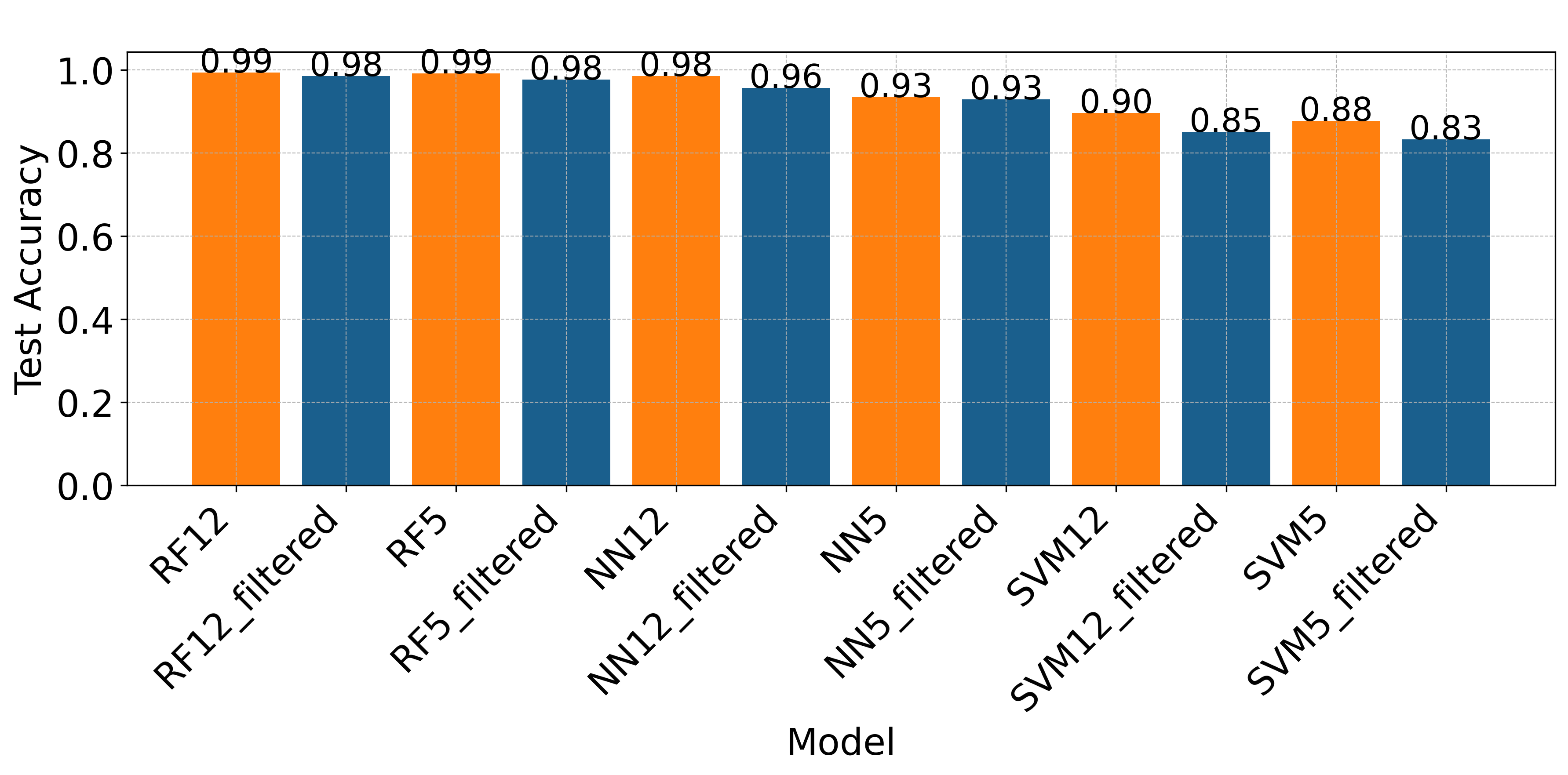}
    \caption{Accuracy Comparison with and without Filtering}
    \label{fig:barchart_acurracy}
\end{figure}

\Cref{tbl:performance} summarizes our experimental results across all model configurations. For visual comparison, we provide the accuracy scores as bar plots in \Cref{fig:barchart_acurracy}. Models are denoted with suffixes `5' or `12' indicating the wavelet decomposition levels used for feature extraction, while the suffix `\_filtered' identifies models trained on datasets where flows with fewer than 20 packets were removed.

\begin{table}[!htp]
    \centering
    \caption{Performance Comparison of ML Models}
    \label{tbl:performance}
    \begin{tabular}{lccc}
        \toprule
        Model & Precision \% & Recall \% & F1-score \% \\
        \midrule
        RF12  & 99 & 99 & 99 \\
        RF12\_filtered  & 98 & 98 & 98 \\
        RF5  & 99 & 99 & 99 \\
        RF5\_filtered  & 98 & 98 & 98 \\
        NN12  & 99 & 98 & 98 \\
        NN12\_filtered & 96 & 96 & 96 \\
        NN5  & 93 & 93 & 93 \\
        NN5\_filtered  & 93 & 93 & 93 \\
        SVM12  & 90 & 90 & 90 \\
        SVM12\_filtered  & 85 & 85 & 85 \\
        SVM5  & 88 & 88 & 88 \\
        SVM5\_filtered  & 83 & 83 & 83 \\
        \bottomrule
    \end{tabular}
\end{table}

\begin{figure*}[ht]
     \centering
     \begin{subfigure}[b]{0.23\textwidth}
         \centering
         \includegraphics[width=1\textwidth]{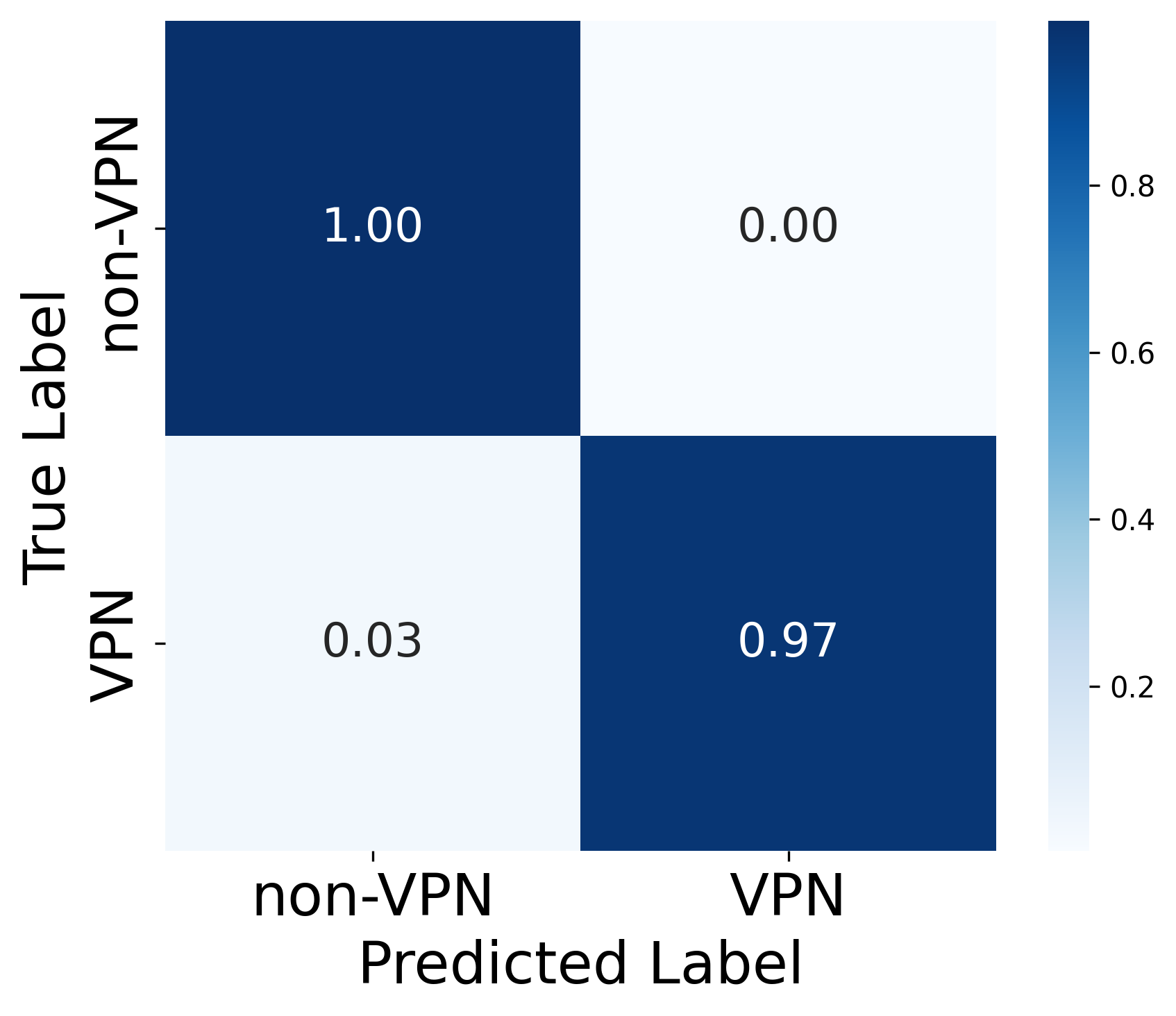}
         \caption{RF12}
     \end{subfigure}
     \hfill
     \begin{subfigure}[b]{0.23\textwidth}
         \centering
         \includegraphics[width=1\textwidth]{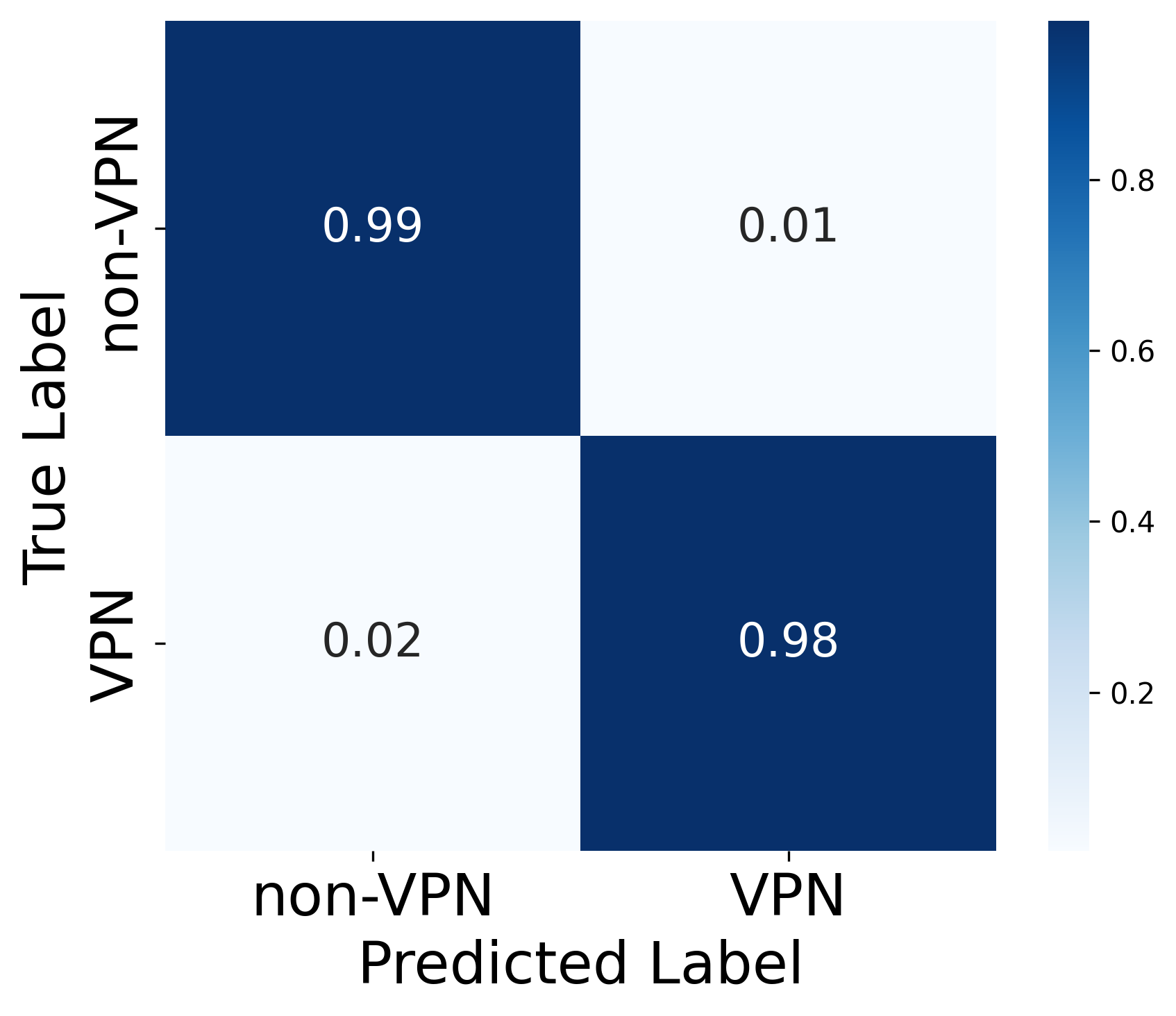}
         \caption{RF12\_filtered}
     \end{subfigure}
     \hfill
     \begin{subfigure}[b]{0.23\textwidth}
         \centering
         \includegraphics[width=1\textwidth]{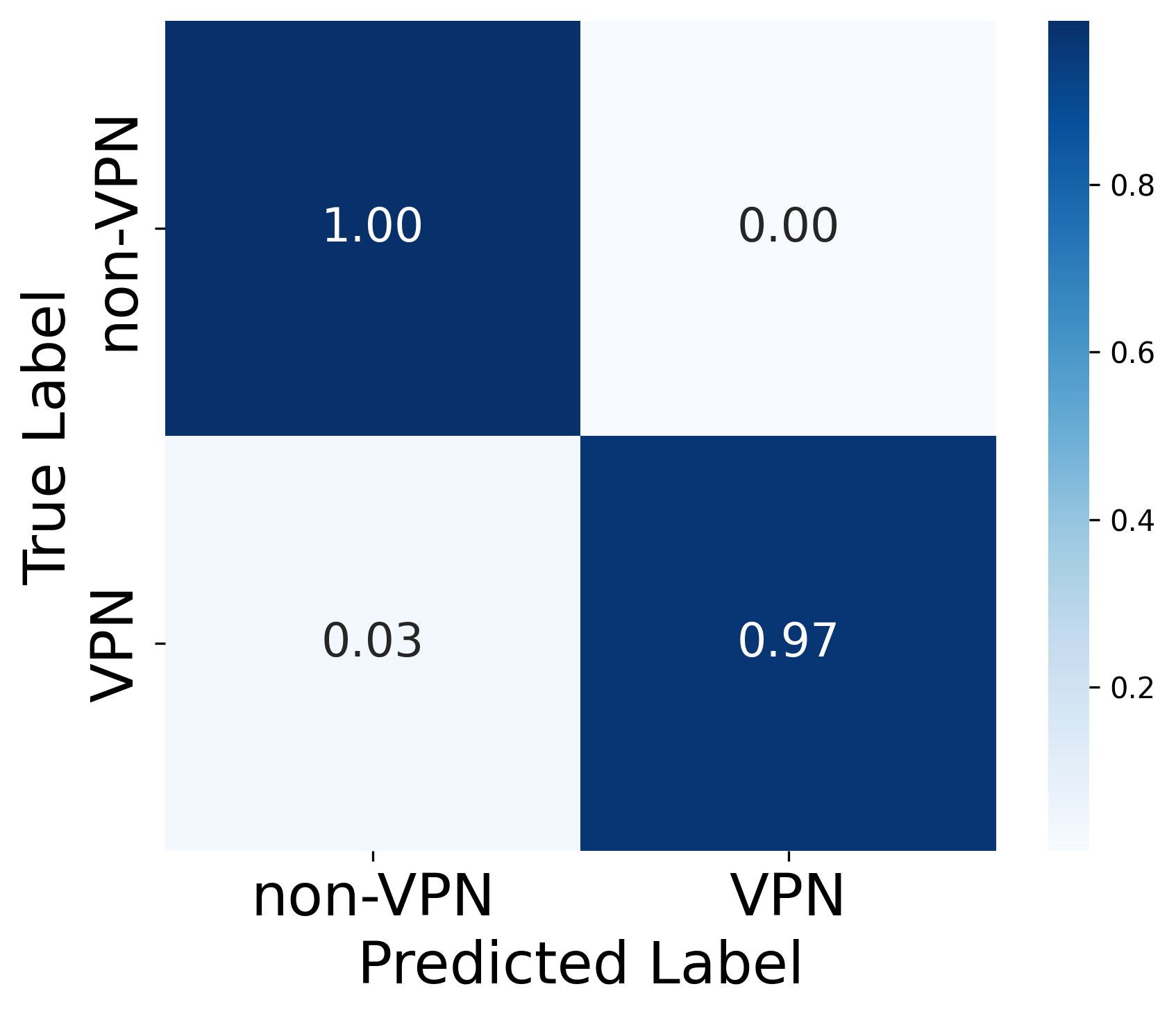}
         \caption{RF5}
     \end{subfigure}
     \hfill
    \begin{subfigure}[b]{0.23\textwidth}
         \centering
         \includegraphics[width=1\textwidth]{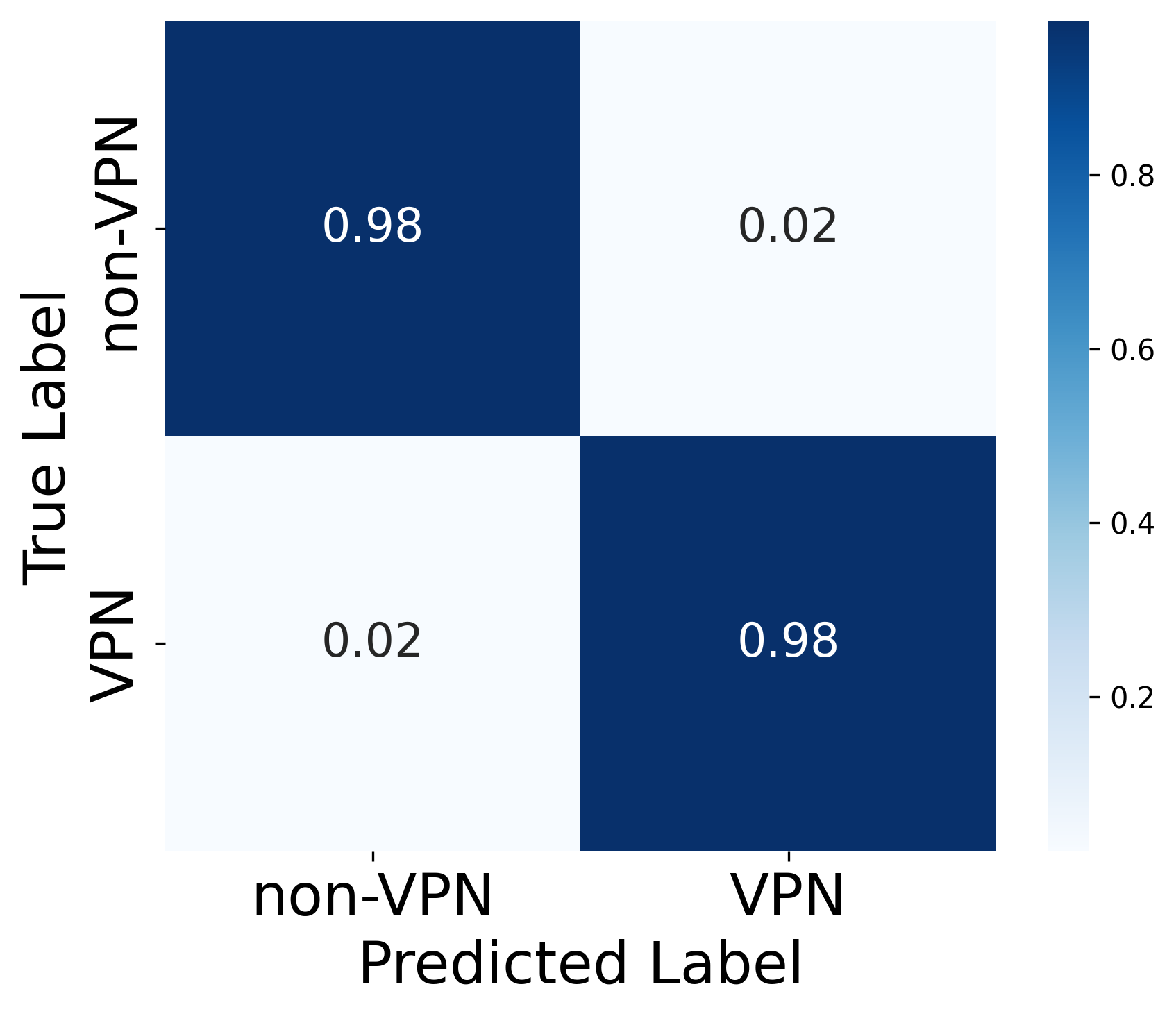}
         \caption{RF5\_filtered}
     \end{subfigure}
      \caption{Confusion Matrices for RF with 5 and 12 Wavelet Decomposition Levels}
        \label{fig:CM}
\end{figure*}

\begin{figure*}[ht]
     \centering
     \begin{subfigure}[b]{0.4\textwidth}
         \centering
         \includegraphics[width=\textwidth,trim=0 0 0 0,clip]{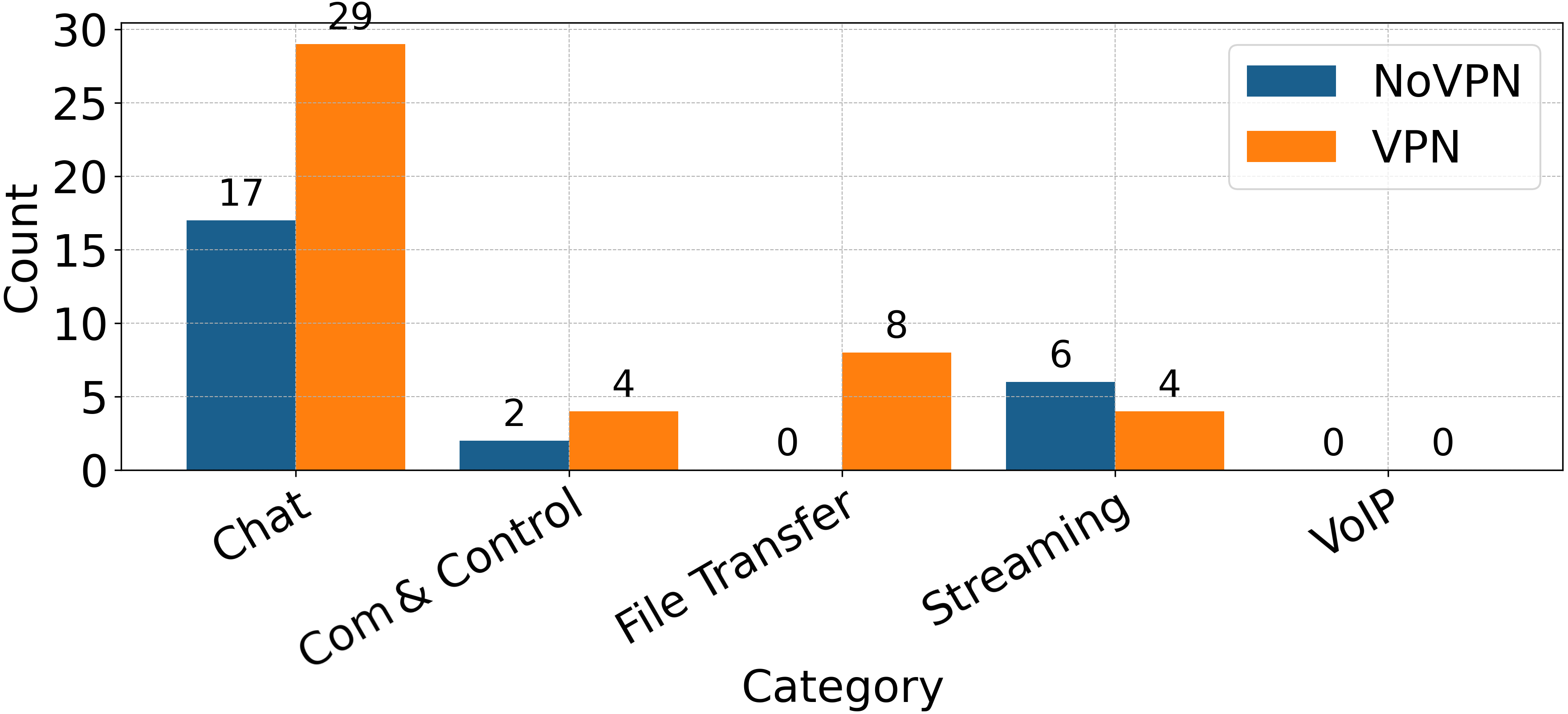}
         \caption{RF12}
     \end{subfigure}
     \hfill
     \begin{subfigure}[b]{0.4\textwidth}
         \centering
         \includegraphics[width=\textwidth,trim=0 0 0 0,clip]{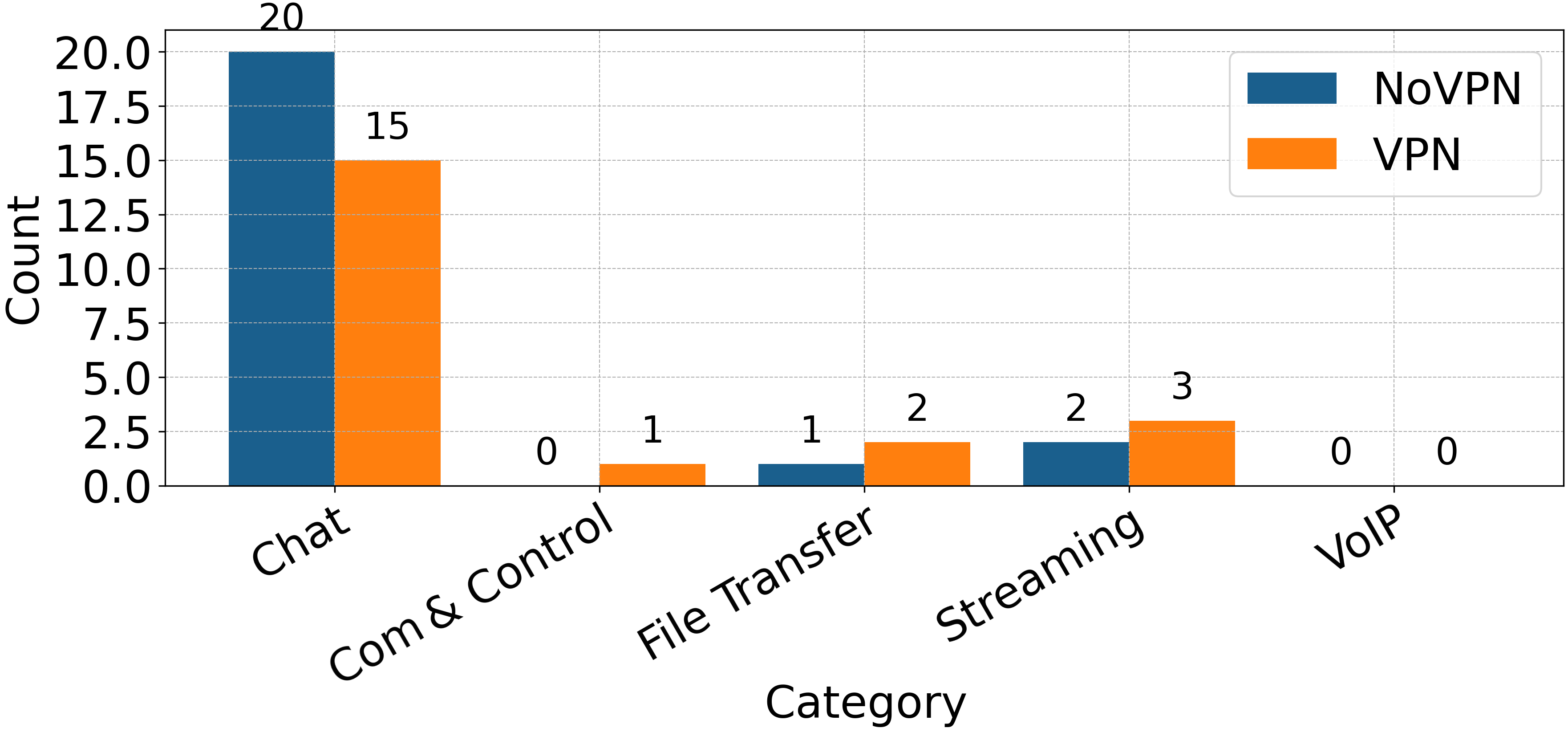}
         \caption{RF12\_filtered}
     \end{subfigure}\\
     \begin{subfigure}[b]{0.4\textwidth}
         \centering
         \includegraphics[width=\textwidth,trim=0 0 0 0,clip]{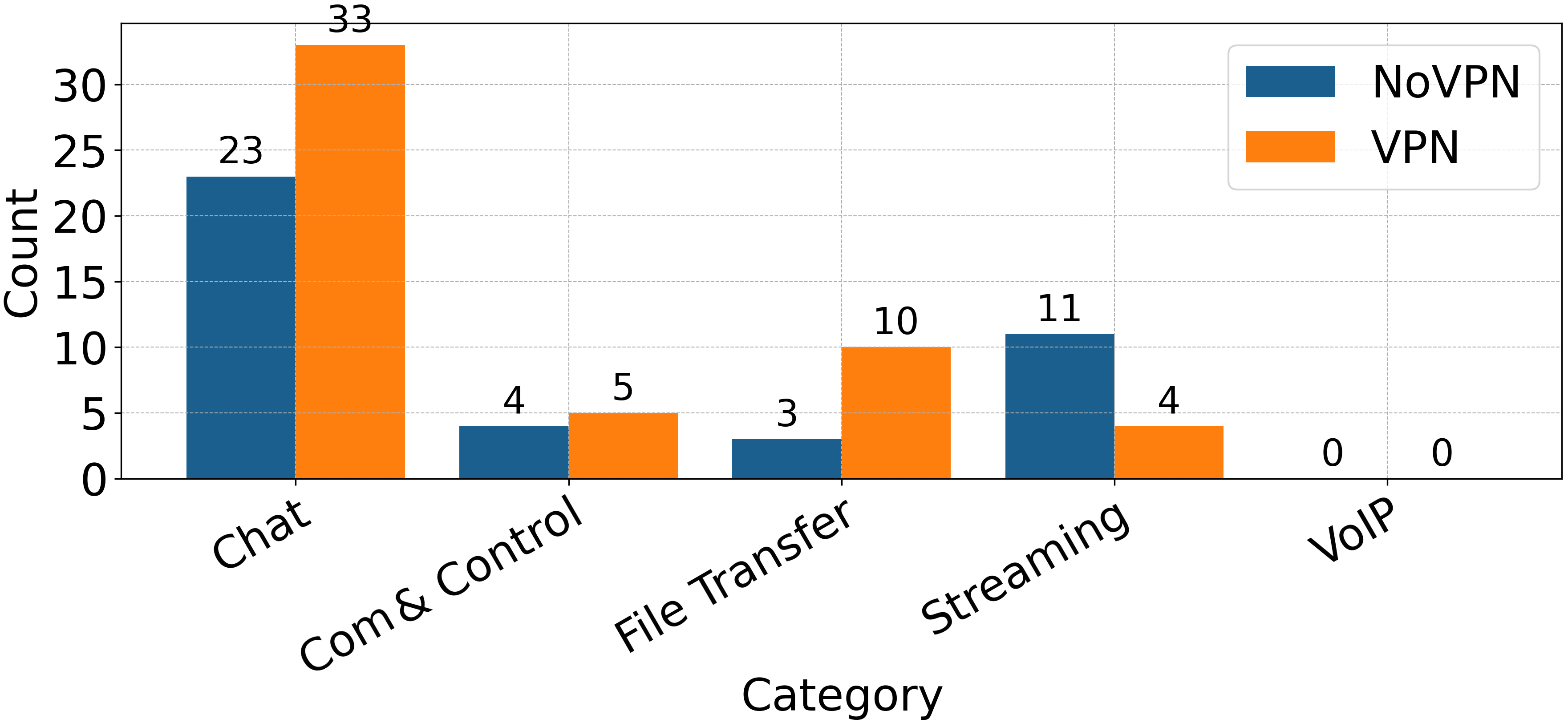}
         \caption{RF5}
     \end{subfigure}
     \hfill
    \begin{subfigure}[b]{0.4\textwidth}
         \centering
         \includegraphics[width=\textwidth,trim=0 0 0 0,clip]{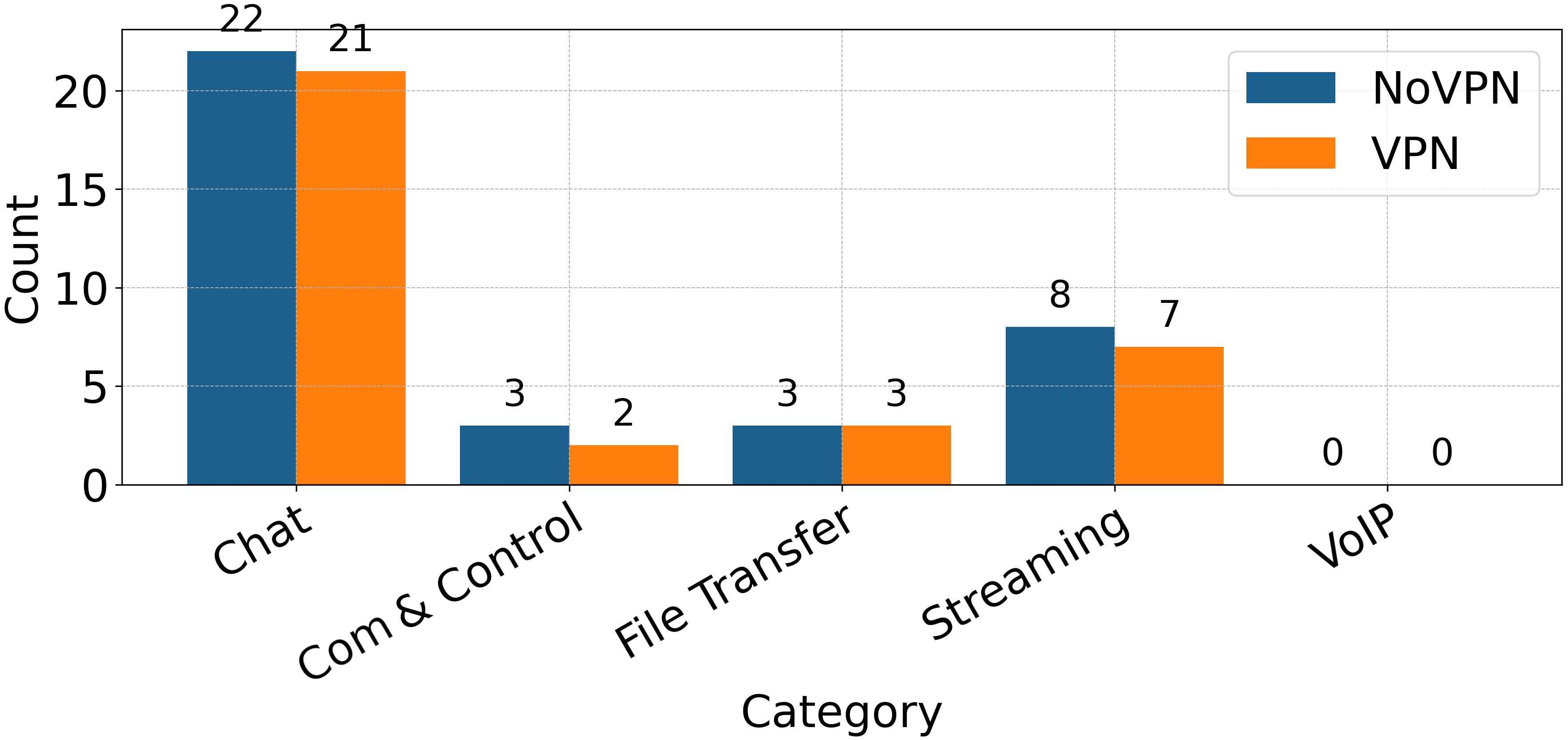}
         \caption{RF5\_filtered}
     \end{subfigure}
     \caption{Distribution of Misclassified Samples Across Different Traffic Categories for RF 5 and 12 levels}
     \label{fig:Misclassification}
\end{figure*}

From \Cref{tbl:performance}, we observe that RF maintains consistently high F1-scores of 99\% at both decomposition levels, with only a minor decrease to 98\% after filtering. This remarkable stability across both preprocessing conditions demonstrates RF's robust feature extraction capability even with minimal data. NNs exhibit sensitivity to decomposition levels, with NN12 achieving an F1-score of 98\%, slightly decreasing to 96\% after filtering, while NN5 remains stable at 93\% across both filtered and unfiltered scenarios. This 5\% performance gap between NN12 and NN5 suggests that NNs benefit from the additional spectral information in higher decomposition levels. SVMs show the most significant impact from filtering, with F1-scores for SVM12 dropping from 90\% to 85\% (5\% decrease), while SVM5 declines from 88\% to 83\% (also 5\% decrease), indicating consistent sensitivity to dataset reduction regardless of decomposition level. This sensitivity likely stems from SVM's reliance on support vectors at the class boundary, which become more sparse after filtering.

The accuracy patterns in \Cref{fig:barchart_acurracy} confirm these observations. RF maintains high stability at both decomposition levels, showing only minimal reductions from 99\% to 98\% after filtering. NN demonstrates moderate sensitivity, with NN12 accuracy decreasing from 98\% to 96\%, while NN5 remains unchanged at 93\% post-filtering. SVM exhibits the most pronounced filtering impact, with SVM12 accuracy declining from 90\% to 85\% and SVM5 dropping from 88\% to 83\%.

Since RF achieved consistently the best performance across both decomposition configurations, we focus our detailed analysis on its confusion matrices and misclassification patterns.

\Cref{fig:CM} illustrates the classification performance across different RF configurations. RF12 achieves 100\% accuracy for non-VPN traffic and 97\% for VPN traffic, with minimal misclassification rates of 0\% and 3\%, respectively. The filtered version maintains strong performance, with 99\% accuracy for non-VPN and 98\% for VPN traffic. 
RF5 exhibits similar performance, achieving 100\% accuracy for non-VPN and 97\% for VPN traffic. After filtering, RF5 shows a small drop in non-VPN classification accuracy to 98\%, while VPN accuracy improved to 98\%. These results confirm RF's robustness, with only minimal variations in misclassification rates across different wavelet levels and filtering conditions.

\Cref{fig:Misclassification} reveals category-specific classification challenges. Chat traffic presents the highest error counts, with RF12 showing 46 total misclassifications (17 non-VPN, 29 VPN) before filtering, reducing to 35 (20 non-VPN, 15 VPN) after filtering. When viewed relative to category size (1301 Chat samples), this represents approximately 3.5\% of Chat samples misclassified. The higher error rate likely stems from Chat's variable traffic patterns and similarities between encrypted and non-encrypted protocols. File Transfer and Command \& Control categories show marked improvement with filtering—RF12's misclassifications in File Transfer drop from 8 to 3 cases, suggesting that longer flows contain more distinctive VPN signatures. RF5 shows similar patterns but with generally higher misclassification counts across all categories, particularly visible in the Chat and Streaming categories.

Notably, as shown in \Cref{fig:Misclassification}, VoIP traffic achieves zero misclassifications in the RF12 model under both filtered and unfiltered conditions, indicating highly distinctive traffic patterns between VPN and non-VPN variants despite the 80\% dataset reduction from filtering. This exceptional performance suggests that VoIP applications maintain unique temporal signatures despite encryption, likely due to their real-time streaming requirements.

When interpreting the reduced misclassification counts after filtering, it's important to consider the dramatic reduction in available samples—up to 95.52\% in File Transfer and 91.64\% in Command \& Control categories, as shown in \Cref{tbl:DS}. Thus, while absolute misclassification counts decrease after filtering, the error rates as a percentage of category size actually increase slightly in several categories (File Transfer error rate increases from 0.05\% to 0.37\% after filtering). This observation aligns with the overall F1-score patterns in \Cref{tbl:performance}, confirming that filtering generally causes a slight performance degradation.

The performance comparison between J=12 and J=5 reveals a consistent advantage for deeper wavelet decomposition, with improvements ranging from 1\% (RF12 vs. RF5) to 5\% (NN12 vs. NN5) in F1-scores. However, these relatively modest gains suggest diminishing returns that may not justify the increased computational overhead in all deployment scenarios, particularly in resource-constrained environments where model efficiency is prioritized alongside accuracy.

\section{Conclusion}
\label{sec:conc}

This study evaluated binary VPN traffic classification using wavelet-based features and machine learning models, focusing on the impact of wavelet decomposition levels and dataset filtering. Our results show that RF consistently achieves the highest classification accuracy, maintaining robustness across different preprocessing conditions, while SVM exhibits the greatest sensitivity to dataset reductions. We found that deeper wavelet decompositions improve classification performance, particularly for variable traffic types, though the marginal gains may not justify the additional computational overhead.
Future research could explore adaptive wavelet decomposition techniques that dynamically adjust to flow characteristics and investigate ensemble approaches combining multiple decomposition levels for enhanced robustness across diverse traffic conditions.

This work utilizes the VNAT dataset~\cite{10044382} as the primary data source. All processed datasets derived from these traces---including extracted network flows, filtered datasets, and wavelet-based feature representations---along with analysis scripts, trained models, and performance metrics necessary to reproduce our results are available at~\cite{github-repo}, ensuring full reproducibility of the machine learning experiments.

\section*{Acknowledgement}

Supported by the János Bolyai Research Scholarship of the Hungarian Academy of Sciences. 
This work was also supported by project TKP2021-NVA-02 (financed under the TKP2021-NVA scheme),
implemented with support provided by the Ministry of Culture and Innovation of Hungary from the National Research, Development and Innovation Fund.

\printbibliography

\end{document}